\documentclass[12pt]{article}
\usepackage{citesort,graphicx}
\newcommand*{\F}{\mathrm{F}}
\renewcommand*{\H}{\mathrm{H}}
\newcommand*{\D}{\mathrm{D}}

\newcommand*{\rd}{\mathrm{d}}
\newcommand*{\rA}{\mathrm{A}}
\newcommand*{\rP}{\mathrm{P}}

\newcommand*{\Ecol}{E_{\mathrm{col}}}
\newcommand*{\Mtr}{M_{\mathrm{tr}}}
\newcommand*{\Inew}{I^{\mathrm{new}}}

\begin{document}
\parindent=0mm
\baselineskip=25pt

\begin{center}
{\Large\bfseries The HF{\boldmath $(v'=3)$} forward scattering peak \\ of the {\boldmath $\F+\H_2$} reaction revisited}

\bigskip\bigskip\bigskip
\baselineskip=20pt

{\large
{\sffamily Lev Yu.\ Rusin\footnote{E-mail:\ \texttt{rusin@chph.ras.ru}},
\ \ Mikhail B. Sevryuk\footnote{E-mail:\ \texttt{sevryuk@mccme.ru}}}

\bigskip

{\slshape Institute of Energy Problems of Chemical Physics, \\ Russia Academy of Sciences, \\ Leninski\`{\i} prospect 38, Bldg.~2, 119334 Moscow, Russia}

\bigskip\medskip

{\sffamily J. Peter Toennies\footnote{E-mail:\ \texttt{jtoenni@gwdg.de}}}

\bigskip

{\slshape Max-Planck-Institut f\"ur Dynamik und Selbstorganisation, \\ Bunsenstrasse 10, \mbox{D-37073} G\"ottingen, Germany}
}

\bigskip\bigskip\bigskip
\baselineskip=17pt

\parbox{14cm}{\small A quantum mechanical coupled-channel scattering calculation on the Stark--Werner potential energy surface is used to study the $\F+\H_2(v=0;j=0,1,2) \to \H+\H\F(v',j')$ reaction at collision energies of 1.84, 2.74, and 3.42 kcal/mol. The dependence of the vibrationally and rotationally resolved differential cross sections $\rd\sigma_{v'j'}/\rd\Omega$ on the product vibrational levels $v'=0$, $1$, $2$, and $3$ as well as on the reactant and product rotational levels is analyzed. The HF$(v'=3)$ center-of-mass forward scattering peak is shown to be caused by the superposition of two effects, namely, the absence of the HF$(v'=3;j')$ products with large $j'$ values due to energy constraints and the growth of the rotationally resolved HF$(v',j')$ forward scattering peak with small $j'$ values as $v'$ increases.}
\end{center}

\newpage
\parindent=5mm
\baselineskip=19pt

\noindent
{\Large\textbf{I. Introduction}}

\bigskip

The $\F+\H_2$ reaction (and its isotopomers $\F+\D_2$ and $\F+\H\D$) has served as one of most important benchmark elementary chemical reactions for at least the last four decades \cite{PP69PT69,PS77And80JBK81Sch85,Man97,YZ08}. During this period, the $\F+\H_2$($\D_2$, HD) interactions have been the subject of extensive studies, both experimental, mainly in crossed molecular beams, and theoretical. The first molecular beam data on the dynamics of these reactions was published by Lee and co-workers in 1970 \cite{SSP70}. In the mid-eighties the same group reported the vibrationally resolved angular distributions of the $\F+\H_2$($\D_2$, HD) reaction products at various collision energies $\Ecol$ \cite{NWR84aNWR84b,NWR85a,NWR85b}.

In their milestone paper \cite{NWR85a}, Neumark et al.\ measured the vibrationally resolved center-of-mass (CM) differential cross sections (DCSs) of the $\F+\H_2(v=0;j=0,1,2)$ reaction at collision energies of $\Ecol=1.84$, 2.74, and 3.42 kcal/mol. The most intriguing observation of Lee and co-workers \cite{NWR84aNWR84b,NWR85a} was an unexpected noticeable forward peak in the angular distribution of the HF$(v'=3)$ product \cite{bibnote}, which increases in intensity with increasing collision energy. The angular distributions of the HF$(v'=2)$ and HF$(v'=1)$ molecules do not exhibit such a peak. A similar but less pronounced peak can be seen in the CM DCSs of the DF$(v'=4)$ product from the $\F+\D_2$ reaction \cite{NWR85b}. The presence of these peaks has been confirmed in subsequent crossed beam experiments for the $\F+\H_2$ reaction \cite{WDQ08} as well as for the $\F+\D_2$ reaction \cite{FMR95FMR96FMR97,FMR98}.

The vibrationally selective forward peaks in the angular distributions of the HF$(v'=3)$ and DF$(v'=4)$ products from the $\F+\H_2$ and $\F+\D_2$ reactions, respectively, were attributed by Neumark et al.\ \cite{NWR84aNWR84b,NWR85a,NWR85b} to quantum mechanical (QM) Feshbach resonances, i.e., metastable states formed on the vibrational adiabatic potentials in the potential energy surface (PES) transition region \cite{YZ08,Fes92,Zhang99,Liu01CS02}. It is interesting to note that some resonances in the $\F+\H_2$ and $\F+\D_2$ reactions were first theoretically predicted three and a half decades ago \cite{WJL73SBK75aSBK75b} in collinear collisions. However, subsequent to their observation these peaks for both the reactions were reproduced in quasiclassical trajectory calculations on various PESs of the $1^2\rA'$ ground state of the $\F\H_2$ system \cite{FMR98,TS88AHN93,ABH94a,ABH94bABH94cABH94dABF96,CHW98,MAB99}. In view of the classical nature of these calculations this was regarded as evidence against an explanation of the peaks by a quantum resonance \cite{Man97,ABH98}. Moreover, Castillo et al.\ \cite{CMS96} concluded, from an analysis of some QM characteristics of the $\F+\H_2$ scattering, that the forward peak in the angular distribution of the HF$(v'=3)$ molecules results from tunneling through the combined centrifugal and potential energy barrier at large values of the total angular momentum $J$ rather than from a resonance. On the other hand, the data of several subsequent studies, both experimental \cite{DLL00} and theoretical \cite{DMC99SCT99SC00,CS00,CS03ACS04ACF05a}, favored the original resonance explanation proposed by Lee and co-workers \cite{NWR84aNWR84b,NWR85a}. This controversy has also been discussed in several reviews \cite{Liu01CS02,AC03Liu06}. Recently, a new crossed molecular beam experiment on the $\F(^2\rP_{3/2})+\H_2(v=j=0)$ reactive scattering was carried out by Wang et al.\ \cite{WDQ08} in the collision energy $\Ecol$ range from 0.4 to 1.2 kcal/mol. Based on a detailed QM simulation they concluded \cite{WDQ08} that the forward peak in the angular distribution of the HF$(v'=3)$ product is generated by a slowing-down during passage over the centrifugal barrier in the exit valley, with a small contribution from a shape resonance \cite{YZ08,Zhang99} at $\Ecol$ slightly above 0.5 kcal/mol, and, moreover, that Feshbach resonances do not contribute to the forward peak.

In crossed beam experiments of 2006, intense forward peaks were discovered in the angular distributions of the HF$(v'=2)$ product from the $\F(^2\rP_{3/2})+\H_2(v=j=0)$ reaction at a collision energy of $\Ecol=0.52$ kcal/mol \cite{QRC06} and the $\F(^2\rP_{3/2})+\H_2(v=0;j=1)$ reaction at a collision energy of $\Ecol=0.19$ kcal/mol \cite{RCQ06}, much lower than sampled by Neumark et al.\ in the eighties \cite{NWR84aNWR84b,NWR85a}. In contrast to the forward peak of the HF$(v'=3)$ molecules discussed above, the peaks pertaining to HF$(v'=2)$ are unanimously regarded as consequences of Feshbach resonances \cite{YZ08,QRC06,RCQ06,YXZ07,Yang07,SSA07SFC07Sok08}. Moreover, in both the reactions $\F+\H_2(v=j=0)$ and $\F+\H_2(v=0;j=1)$, the forward peak in the angular distribution of the HF$(v'=2)$ product is caused by two dynamical resonances which interfere with each other.

Numerous convincing manifestations of QM Feshbach resonances have also been found at integral and differential cross sections of the $\F+\H\D \to \D+\H\F$ reaction at various collision energies. The resonances in this reaction have been extensively studied both experimentally and theoretically since 2000 \cite{DLL00,SSM00aSSM00b}, as discussed in the reviews \cite{YZ08,Liu01CS02,AC03Liu06,Yang07,LSM02} and references therein. In the last four years several additional important publications have appeared \cite{LWL07,FCA07,FAC08,RCQ08YLC09Alt10DXW10}. The $\F+\H\D$ interaction is interesting from the viewpoint of its stereodynamics \cite{LZS09,SFC09}.

In this article, we reexamine the controversial role of resonances in the forward scattering of the HF$(v'=3)$ product from the $\F+\H_2$ reaction employing an approach first described in our preceding publications in Russian \cite{RST03RST07}. Our QM simulations of the $\F+\H_2$ interaction at the collision energies of the experiment by Lee and co-workers \cite{NWR85a} ($\Ecol=1.84$, 2.74, and 3.42 kcal/mol) explore the trends in the behavior of the vibrationally and rotationally resolved DCSs of the HF$(v',j')$ scattering as the vibrational quantum number $v'$ increases from $0$ to $3$. Although the cross sections for the formation of the HF product from the $\F+\H_2$ reaction in the ground vibrational state are very small, HF$(v'=0)$ molecules have been detected in a crossed molecular beam experiment \cite{RT00}. Our analysis shows that the vibrationally specific forward peak of the HF$(v'=3)$ product can be explained by the superposition of two independent effects which reinforce each other. One of these effects is a purely energy restriction, while the other one touches upon \emph{all} the vibrational states $v'$ of HF molecules and is therefore hardly of mainly resonance nature. Thus, this study provides additional evidence in support of the conclusions of Wang et al.\ \cite{WDQ08} against a resonance origin of the peak in question.

The paper is organized as follows. After a brief discussion of the calculation details in section~II, we analyze the QM vibrationally and rotationally resolved DCSs of the $\F+\H_2(v=0;j=0,1,2)$ reaction in section~III\@. The remarks of section~IV conclude the article.

\bigskip\medskip

\noindent
{\Large\textbf{II. DCS Calculations}}

\bigskip

The ab initio Stark--Werner (SW) PES \cite{SW96} of the $1^2\rA'$ ground state of the $\F\H_2$ system has had an enormous impact on the theoretical development since 1993 \cite{Man97,MSW93}. Considerable effort has since been devoted to further improving this surface in the entrance valley which has led to the HSW PES \cite{CHW98,HW97}, \mbox{SW-LR} PES \cite{ACF02}, \mbox{SW-LR-SO} PES \cite{ACF02}, and PES III \cite{ACF03ACF05b}, or in the exit valley (the SWMHS PES \cite{HGM05}), or in both the valleys simultaneously (the PES IV \cite{FAC08}). In addition, two totally new ab initio surfaces XXZ \cite{QRC06,XXZ06} and FXZ \cite{FXZ08} were recently produced. The calculations of the present paper are carried out on the standard SW PES \cite{SW96} to facilitate comparison with the data of the previous publications \cite{Man97,CHW98,CMS96,CS00,ACF02,TA04}. The SW surface is still widely used in studies of the $\F+\H_2$($\D_2$, HD) reactions \cite{LZS09,AAK09,XC09,Wang10}. Moreover, our observations are of a qualitative character, and we have not attempted a direct comparison with the experiment. Recall, however, that the experimental forward peak in the angular distributions of the HF$(v'=3)$ product from the $\F+\H_2$ reaction is less pronounced than that obtained in QM simulations on the SW PES \cite{Man97,CHW98,CMS96,ACF02,TA04}. The same situation holds for the forward peak in the angular distributions of the DF$(v'=4)$ product from the $\F+\D_2$ reaction \cite{MAB99}.

As in most of the previous theories of the $\F+\H_2$($\D_2$, HD) reactive scattering \cite{Man97}, contributions from the excited state F${}^{\ast}(^2\rP_{1/2})$ of the fluorine atom reactant were neglected. This approximation is fully justified for collision energies $\Ecol\geq 1.84$ kcal/mol in the context of the present study \cite{LWL07,TA04,AMW00,RST05LAL08}. Also according to Tzeng and Alexander \cite{TA04}, these contributions affect mainly backward scattering. Note, however, that several ab initio multi-state surfaces for all of the three lowest electronic states $1^2\rA'$, ${}^2\rA''$, and $2^2\rA'$ of the $\F\H_2$ system as well as for the $1^2\rA'\leftrightarrow 2^2\rA'$ nonadiabatic coupling have been constructed: the ASW \cite{AMW00}, MASW \cite{XZH04}, \mbox{LWA-5} \cite{LWL07}, and \mbox{LWA-78} \cite{LWL07} PESs.

For each collision energy $\Ecol=1.84$, 2.74, and 3.42 kcal/mol \cite{NWR85a}, the vibrationally and rotationally resolved DCSs $\rd\sigma_{v'j'}/\rd\Omega$ of the $\F+\H_2(v=0;j) \to \H+\H\F(v',j')$ reactions were calculated for $j=0$, $1$, and $2$ using the ABC program \cite{SCM00}. This code solves the Schr\"odinger equation for the motion of the three nuclei on a given PES by a coupled-channel method in the Delves hyperspherical coordinates \cite{PP87PP89}. The calculations were performed for the total angular momentum $J$ ranging from $0$ to a maximum value $J_{\max}$. For $\Ecol=1.84$ kcal/mol, the convergence parameters of the ABC program were set at $J_{\max}=25$, $E_{\max}=1.7$ eV, $j_{\max}=17$, $k_{\max}=4$, $\rho_{\max}=12$ bohr, and $\Mtr=150$, and for $\Ecol=2.74$ and 3.42 kcal/mol at $J_{\max}=30$, $E_{\max}=2.5$ eV, $j_{\max}=21$, $k_{\max}=5$, $\rho_{\max}=12$ bohr, and $\Mtr=200$. Here $E_{\max}$ is the maximum internal energy of the $\H_2$ reagent and HF product admissible in the basis functions, $j_{\max}$ is the maximum rotational quantum number of the $\H_2$ reagent and HF product, $k_{\max}$ is the maximum (in absolute value) helicity quantum number of the $\H_2$ reagent and HF product, $\rho_{\max}$ is the maximum hyperradius $\rho$ of the system used while solving the hyperradial coupled-channel equations, and $\Mtr$ is the number of propagation sectors involved in solving those equations (for details, see the paper by Skouteris et al.\ \cite{SCM00}). Some test calculations with other values of the convergence parameters indicate that the chosen values of the parameters yield the DCSs for almost all the HF$(v',j')$ states with a relative accuracy of better than $\sim 1$\%.

\bigskip\medskip

\noindent
{\Large\textbf{III. Results and Discussion}}

\bigskip

\noindent
{\large\textbf{a.\ Forward Scattering Coefficients}}

\bigskip

As an example, Figure~1 presents the rotationally unresolved DCSs $\rd\sigma_{v'}/\rd\Omega$ and some rotationally resolved DCSs $\rd\sigma_{v'j'}/\rd\Omega$ of the $\F+\H_2(v=0;j=2)$ reaction at a collision energy of $\Ecol=3.42$ kcal/mol. As a whole, the DCSs depend on $\Ecol$ only rather weakly. There is, however, a distinct shift towards the smaller $\theta$ values as the collision energy increases. In the second panel from the bottom of Figure~1, one observes a moderate forward peak in the angular distribution of the HF$(v'=2)$ molecules. On the other hand, the peak in the angular distribution of the HF$(v'=3)$ molecules in the region of small $\theta\leq 20^{\circ}$ is very strongly pronounced in the lowest panel of Figure~1. For other values of collision energy $\Ecol$ and initial rotational quantum numbers $j$ of the target $\H_2$ molecule, this peak is also largely confined to the CM scattering angle range $0^{\circ}\leq\theta\leq 20^{\circ}$ in agreement with the results of the previous QM simulations on the SW PES \cite{Man97,CHW98,CMS96,ACF02,TA04}. Consequently, to characterize the relative contributions from various $j'$ values to the forward peak, each $(v',j')$ state of the HF product from the $\F+\H_2(v=0;j)$ reaction will be assigned the quantity
\begin{equation}
I_{v'j'}=100
\frac{\displaystyle \int_{0^{\circ}}^{20^{\circ}}
\frac{\rd\sigma_{v'j'}}{\rd\Omega}(\theta)\,\rd\theta -
\int_{20^{\circ}}^{40^{\circ}}
\frac{\rd\sigma_{v'j'}}{\rd\Omega}(\theta)\,\rd\theta}
{\displaystyle \int_{0^{\circ}}^{180^{\circ}}
\frac{\rd\sigma_{v'j'}}{\rd\Omega}(\theta)\,\rd\theta},
\label{eq1}
\end{equation}
which we call the \emph{forward scattering coefficient} of the HF$(v',j')$ molecules. This coefficient can be either positive or negative. The larger the positive $I_{v'j'}$ coefficient, the more pronounced the forward peak in the HF$(v',j')$ angular distribution. However, since in many cases the rotationally resolved DCSs of the $\F+\H_2(v=0;j)$ reaction oscillate as $\theta\to 0$ (see Figure~1), it is more appropriate to speak here of an ``average'' forward angular distribution. The forward scattering coefficients of eq~\ref{eq1} involve the integrals $\int(\rd\sigma_{v'j'}/\rd\Omega)(\theta)\,\rd\theta$ instead of the expressions $\int(\rd\sigma_{v'j'}/\rd\Omega)(\theta)\sin\theta\,\rd\theta$ since otherwise the DCS behavior at small angles $\theta$ would be suppressed. The integrals in eq~\ref{eq1} just compare the mean values of the DCSs over the corresponding angular ranges and do not measure the scattering into these angular ranges. The values of the $I_{v'j'}$ coefficients for various $\Ecol$, $j$, $v'$, and $j'$ are presented in Figure~2.

For $v'=0$, all the $I_{0j'}$ coefficients are tiny for all $\Ecol$ and $j$ except for $I_{0j'}$ with $j'\geq 14$ at $\Ecol=3.42$ kcal/mol and $j=1$. The same is also essentially true for $v'=1$, but the $I_{1j'}$ coefficients for $j'\leq 5$, although very small, are noticeably larger on the whole than the $I_{1j'}$ coefficients for $j'\geq 6$. For $v'=2$, many of the $I_{2j'}$ coefficients for $j'\leq 5$ are already rather large, whereas most of the $I_{2j'}$ coefficients for $j'\geq 6$ remain small. Finally, for $v'=3$, conservation of energy restricts the rotational levels of the HF$(v'=3)$ molecules. The corresponding $I_{3j'}$ coefficients for $j'\leq 5$ are on the whole much larger than the $I_{2j'}$ coefficients. Note that the smaller $j$ and the higher $\Ecol$ are, the larger are the $I_{3j'}$ coefficients for any fixed $j'\leq 2$. The same trends persist, in general, for $j'\geq 3$. For this reason, lower initial rotational excitations of the target $\H_2$ molecule and higher collision energies lead to a more strongly pronounced forward peak of $\rd\sigma_3/\rd\Omega$.

The forward scattering coefficients $I_{v'j'}$ compiled in Figure~2 provide the following explanation for the origin of the forward peak of the HF$(v'=3)$ product. As the vibrational quantum number of the HF molecules increases from $v'=0$ to $v'=3$, the $I_{v'j'}$ coefficients grow rapidly for small $j'$ ($j'\leq 4$ or $j'\leq 5$), whereas for larger $j'$ they do not tend to increase and remain small for all $v'$. The fact that the forward scattering peak of the $\rd\sigma_{v'}/\rd\Omega$ DCSs is present for $v'=3$ and absent for $v'\leq 2$ is caused by the joint influence of two effects which are independent but nonetheless strengthen each other.

First we note that among the rotationally resolved DCSs $\rd\sigma_{v'j'}/\rd\Omega$, only the cross sections corresponding to small $j'$ values possess such a peak. However, for $v'\leq 2$, the HF$(v',j')$ products with small $j'$ constitute only a fraction of all the HF$(v')$ molecules in the given vibrational state $v'$ \cite{PW72PP76CBN98}. Consequently, the presence of the forward peak of the $\rd\sigma_{v'j'}/\rd\Omega$ DCSs with small $j'$ values cannot substantially affect the total angular distribution of the HF$(v')$ molecules. For $v'=3$, on the other hand, as pointed out above, all the HF$(v',j')$ products have small rotational quantum numbers $j'$ due to energy restrictions and exhibit forward scattering peaks. After the summation over $j'$, these peaks yield a forward peak in the rotationally unresolved angular distribution of the HF$(v'=3)$ molecules.

Secondly, as $v'$ increases, the stronger are the forward peaks of the rotationally resolved DCSs $\rd\sigma_{v'j'}/\rd\Omega$ for small $j'$. For this reason, even the sum
\[
\sum_{j'\leq 5}\frac{\rd\sigma_{v'j'}}{\rd\Omega}(\theta),
\]
with all the rotational levels $j'\geq 6$ excluded, exhibits for $v'=3$ a much more pronounced forward peak than for $v'=2$, not to mention $v'\leq 1$.

Thus, the forward peak of the HF$(v'=3)$ molecules discovered experimentally by Neumark et al.\ \cite{NWR84aNWR84b,NWR85a} is most likely explained by the different behaviors of the HF DCSs not only with respect to the vibrational quantum number $v'$ but also with respect to the rotational quantum number $j'$. The first of the effects indicated above, the absence of the HF$(v'=3;j')$ products with large $j'$, is just an energy restriction. The second effect, an increase in the forward scattering peaks of the rotationally resolved DCSs $\rd\sigma_{v'j'}/\rd\Omega$ for small $j'$ as $v'$ grows, is a trend that affects \emph{all} the vibrational states $v'$ of the HF product. The resonance origin of this effect seems therefore rather questionable. We conclude that the analysis of the forward scattering coefficients $I_{v'j'}$ provides evidence against the resonance nature of the forward peak of the HF$(v'=3)$ molecules.

For the rotationally resolved DCSs $\rd\sigma_{v'j'}/\rd\Omega$ the first effect is of no consequence compared to the rotationally unresolved DCSs $\rd\sigma_{v'}/\rd\Omega$. The increase in forward scattering of the rotationally resolved HF products in passing from $v'=2$ to $v'=3$ is therefore much less sharp than that of the rotationally unresolved products as illustrated in Figure~1. The rotationally unresolved DCSs $\rd\sigma_{v'}/\rd\Omega$ (solid black curves in Figure~1) have a forward peak for the $\rd\sigma_3/\rd\Omega$ cross section and with the exception of a small rise in the $\rd\sigma_2/\rd\Omega$ cross section no such peak for $v'\leq 2$. On the other hand, the rotationally resolved DCSs $\rd\sigma_{v'j'}/\rd\Omega$ show a distinct gradual increase in the forward peak for small $j'$ values in passing from $v'=0$ to $v'=1$, $v'=2$, and, finally, to $v'=3$.

A similar situation holds for the $\F+\D_2$ reaction. Figures~\mbox{2--4} in the paper by Mart\'{\i}nez-Haya et al.\ \cite{MAB99} present some rotationally resolved DCSs $\rd\sigma_{v'j'}/\rd\Omega$ of the $\F+\D_2(v=0;j=0,1,2)$ reaction for $v'=2$, $3$, and $4$ at the collision energies of the G\"ottingen experiments \cite{FMR98,BFM98} $\Ecol=2.08$, 3.23, 4.15, and 5.53 kcal/mol. These DCSs, averaged over $j$ according to the experimental rotational distributions of the target $\D_2$ molecules, were obtained from quasiclassical trajectory as well as from quantum mechanical close-coupling calculations, both on the SW PES. For the QM cross sections at $\Ecol\geq 3.23$ kcal/mol and for the quasiclassical cross sections at $\Ecol\geq 4.15$ kcal/mol one finds the same trend as in Figure~1 of the present work, i.e., a relative increase in the forward peaks of the $\rd\sigma_{v'j'}/\rd\Omega$ DCSs in passing not only from $v'=3$ to $v'=4$ but also from $v'=2$ to $v'=3$. Unfortunately, Mart\'{\i}nez-Haya et al.\ \cite{MAB99} did not pay due attention to this phenomenon. The increase in forward peaks in the angular distributions of the DF$(v',j')$ molecules in passing from $v'=3$ to $v'=4$ is, however, much sharper than in the angular distributions of the HF$(v',j')$ molecules in passing from $v'=2$ to $v'=3$.

The gradual evolution of the forward scattering coefficients as $v'$ increases is enhanced if the definition of these coefficients is changed so that the backward scattering of the HF products, which is very strong for $v'\leq 2$ (see Figure~1), is removed. This is achieved by replacing the integration over $\rd\theta$ from $0^{\circ}$ to $180^{\circ}$ in the denominator of the right-hand side of eq~\ref{eq1} by an integration from $0^{\circ}$ to $90^{\circ}$:
\begin{equation}
\Inew_{v'j'}=100
\frac{\displaystyle \int_{0^{\circ}}^{20^{\circ}}
\frac{\rd\sigma_{v'j'}}{\rd\Omega}(\theta)\,\rd\theta -
\int_{20^{\circ}}^{40^{\circ}}
\frac{\rd\sigma_{v'j'}}{\rd\Omega}(\theta)\,\rd\theta}
{\displaystyle \int_{0^{\circ}}^{90^{\circ}}
\frac{\rd\sigma_{v'j'}}{\rd\Omega}(\theta)\,\rd\theta}.
\label{eq2}
\end{equation}
These modified forward scattering coefficients are presented in Figure~3. Their behavior differs but slightly, as a whole, from the behavior of the coefficients of eq~\ref{eq1}, but the absolute values are considerably larger for $v'\leq 2$. Compared with the $I_{v'j'}$ coefficients, the increase in the coefficients of eq~\ref{eq2} for small $j'$ levels in passing from $v'=1$ to $v'=2$ and from $v'=2$ to $v'=3$ is much weaker.

Note that the forward peaks of the \emph{experimental} DCSs for the HF molecules with the maximum vibrational quantum number $v'=3$ are confined to the interval $0^{\circ}\leq\theta\leq 30^{\circ}$ \cite{NWR85a}.

\bigskip\smallskip

\noindent
{\large\textbf{b.\ Contributions from Separate Rotational States}}

\bigskip

To gain further insight into the forward peak in the angular distribution of the HF$(v'=3)$ molecules, we also define the quantities
\begin{equation}
D_{v'j'}=100
\frac{\displaystyle \int_{0^{\circ}}^{20^{\circ}}
\frac{\rd\sigma_{v'j'}}{\rd\Omega}(\theta)\sin\theta\,\rd\theta}
{\displaystyle \int_{0^{\circ}}^{20^{\circ}}
\frac{\rd\sigma_{v'}}{\rd\Omega}(\theta)\sin\theta\,\rd\theta},
\label{eq3}
\end{equation}
so that
\[
\sum_{j'}D_{v'j'}=100
\]
for any $v'$. The $D_{v'j'}$ ratios characterize explicitly the percentage contributions of the $j'$ rotational levels to scattering of the HF$(v')$ product into the angular range $\theta\leq 20^{\circ}$. The values of the quantities of eq~\ref{eq3} for various $\Ecol$, $j$, $v'$, and $j'$ are presented in Figure~4.

As is seen in Figure~4, the forward scattering of the HF$(v')$ molecules is either rotationally hot or ``neutral'' for $v'=0$ or rotationally cold for $v'\geq 1$. The larger the $v'$ is, the smaller are the rotational quantum numbers $j'$ of the HF molecules which make a major contribution to the HF$(v')$ forward scattering. For $v'\geq 2$, this trend still enhances the role of the HF$(v',j')$ molecules with $j'\leq 5$. The dominant contribution to the forward peak of the $\rd\sigma_3/\rd\Omega$ DCSs comes from the HF$(v'=3;j')$ products with very small $j'$ values, namely, with $j'\leq 3$. The only exception is the $\F+\H_2(v=0;j=2)$ reaction at collision energies $\Ecol\geq 2.74$ kcal/mol, for which the $j'=4$ must also be accounted for. This result is in agreement with the data from previous simulations \cite{ABH94a,CHW98}. Note that the forward peak in the angular distribution of the DF$(v'=4)$ product from the $\F+\D_2$ reaction is also rotationally cold \cite{FMR98,MAB99}.

The calculations show that the quantities of eq~\ref{eq3} would not be significantly affected on the whole if the factor $\sin\theta$ is removed from the integrand in both the numerator and denominator.

\bigskip\smallskip

\noindent
{\large\textbf{c.\ DCS Oscillations}}

\bigskip

The oscillations of the DCSs of the HF$(v',j')$ forward scattering are clearly seen in Figure~1. They are especially pronounced for $j'=0$, and for $v'\geq 2$ their amplitudes increase sharply as $\theta\to 0$. The angular ``periods'' of the oscillations lie between $8^{\circ}$ and $20^{\circ}$ in typical cases. These oscillations can be attributed to the oscillations at large values of the total angular momentum $J$ of the reduced $d^J_{k'k}$ entries of the Wigner rotation matrix \cite{Zhang99,Zare88}. Here $k$ is the helicity quantum number of the $\H_2$ reagent while $k'$ is the helicity quantum number of the HF product \cite{Zhang99,Mil69ZM89}. This oscillatory structure is only present in forward and sometimes sideways HF scattering. In backward scattering the major contribution comes from small $J$ values for which the $d^J_{k'k}(\pi-\theta)$ are non-oscillatory functions. Oscillations of the rotationally resolved angular distributions of the HF$(v',j')$ products from the $\F+\H_2$ reaction have been reported previously in theoretical papers \cite{WDQ08,CHW98,SSA07SFC07Sok08,RST03RST07} but have as far as we are aware not been observed experimentally.

The oscillatory structure can be described quantitatively as follows. Let $M_1, M_2, \ldots$ denote the values of the successive maxima of the function $(\rd\sigma_{v'j'}/\rd\Omega)(\theta)$ as $\theta$ increases from $0^{\circ}$ to $40^{\circ}$ and $m_1, m_2, \ldots$ the values of the successive minima. Forward scattering at $\theta=0^{\circ}$ is regarded as a maximum point if
\[
\frac{\rd\sigma_{v'j'}}{\rd\Omega}(0^{\circ}) >
\frac{\rd\sigma_{v'j'}}{\rd\Omega}(0.5^{\circ})
\]
and as a minimum point if the opposite inequality
\[
\frac{\rd\sigma_{v'j'}}{\rd\Omega}(0^{\circ}) <
\frac{\rd\sigma_{v'j'}}{\rd\Omega}(0.5^{\circ})
\]
holds. In the first case $(\rd\sigma_{v'j'}/\rd\Omega)(0^{\circ})=M_1$ and in the second case $(\rd\sigma_{v'j'}/\rd\Omega)(0^{\circ})=m_1$. The choice of $0.5^{\circ}$ in the above inequalities is arbitrary and any other very small positive angle could be used. The quantity
\begin{equation}
A_{v'j'}=100
\frac{\displaystyle \max\{M_1,M_2\}-\min\{m_1,m_2\}}
{\displaystyle \max_{0^{\circ}\leq\theta\leq 90^{\circ}}
\frac{\rd\sigma_{v'j'}}{\rd\Omega}(\theta)}
\label{eq4}
\end{equation}
will be called the \emph{relative oscillation amplitude} for the DCSs of the HF$(v',j')$ forward scattering. If the $\rd\sigma_{v'j'}/\rd\Omega$ cross section has less than two maxima or less than two minima in the interval $0^{\circ}\leq\theta\leq 40^{\circ}$, oscillations are assumed to be absent and we set $A_{v'j'}=0$. The values of the relative oscillation amplitudes $A_{v'j'}$ for various $\Ecol$, $j$, $v'$, and $j'$ are presented in Figure~5.

Figure~5 shows a distinct increase in $A_{v'j'}$ for small $j'\leq 5$ as $v'$ grows from $0$ to $3$. Moreover, the relative amplitudes $A_{v'j'}$ for $v'=1$ and $v'=2$ decrease in general as $j'$ increases, and the values of these amplitudes for the $\F+\H_2(v=0;j=2)$ reaction are smaller, as a whole, than those for the $\F+\H_2(v=0;j)$ reactions with $j=0$ and $j=1$. The reason is that the HF$(v',j')$ products are formed only in collisions with helicity quantum numbers $|k|\leq j$ and $|k'|\leq j'$. Since the oscillations of $d^J_{k'k}$ functions with different $k$ and $k'$ damp each other, the oscillations of the HF$(v',j')$ angular distributions are in general more pronounced for the smaller $j$ and $j'$. Why some $\rd\sigma_{v'j'}/\rd\Omega$ DCSs with $j'=1$ exhibit no oscillations is at present not clear.

If the maximum over the interval $0^{\circ}\leq\theta\leq 90^{\circ}$ in the denominator of the right-hand side of eq~\ref{eq4} is replaced with the maximum over the interval $0^{\circ}\leq\theta\leq 180^{\circ}$, the general behavior of the relative oscillation amplitudes would change only slightly, similarly as in the case of the forward scattering coefficients (eqs~\ref{eq1} and~\ref{eq2}). However, for $v'\leq 2$ and especially for $v'\leq 1$, this replacement considerably reduces many of the relative amplitudes.

\bigskip\smallskip

\noindent
{\large\textbf{d.\ Partial Wave Analysis}}

\bigskip

To examine one more facet of the HF$(v',j')$ angular distributions, we introduce another quantity denoted by $X_{v'j'}^J$, which is the partial DCS of the HF$(v',j')$ product scattering (at fixed $\Ecol$ and $j$) obtained by taking into account the total angular momenta from $0$ to some value $J\leq J_{\max}$. Since the $\rd\sigma_{v'j'}/\rd\Omega$ cross sections are calculated using the total angular momenta from $0$ to $J_{\max}$, we have $X_{v'j'}^{J_{\max}} \equiv \rd\sigma_{v'j'}/\rd\Omega$. The contribution to the $(\rd\sigma_{v'j'}/\rd\Omega)(\theta)$ DCS from a single partial wave corresponding to a given value $J$ of the total angular momentum can then be defined as the difference $X_{v'j'}^J(\theta)-X_{v'j'}^{J-1}(\theta)$ \cite{WDQ08,CMS96,AAK09}, where for $J=0$ we set $X_{v'j'}^{-1}\equiv 0$. Because of interference effects, such a contribution can be negative for some angles $\theta$. The total contribution from the $J$ partial wave to the HF$(v',j')$ forward scattering can be measured by the integral
\[
Y_{v'j'}^J=\int_{0^{\circ}}^{20^{\circ}}
\bigl[ X_{v'j'}^J(\theta)-X_{v'j'}^{J-1}(\theta) \bigr]\,\rd\theta.
\]
For any set of values of $\Ecol$, $j$, $v'$, and $j'$, a small fraction of negative $Y_{v'j'}^J$ integrals among all the nonzero numbers $Y_{v'j'}^0$, $Y_{v'j'}^1$, \ldots\ indicates mainly constructive interference of the partial waves in forward scattering of the HF$(v',j')$ product. In the region $\theta\leq 20^{\circ}$, almost every partial wave reinforces the sum of waves corresponding to smaller momenta $J$. Conversely, a large fraction means mainly destructive interference. In the region $\theta\leq 20^{\circ}$, many of the partial waves attenuate the sums of waves corresponding to the previous values of the total angular momentum $J$.

Our calculations show that for $v'=0$, the fraction of negative $Y_{v'j'}^J$ quantities lies between 20\% and 60\% for most of the values of $\Ecol$, $j$, and $j'$. For $v'=1$ and $v'=2$, it typically lies between 20\% and 50\% and between 10\% and 45\%, respectively. For $v'=3$, this fraction does not exceed 35\% in all the cases and does not exceed 25\% for $j'\leq 4$. Moreover, for many of the sets $(\Ecol,j,j')$, all the integrals $Y_{3j'}^J$ are non-negative. The total fraction of negative $Y_{v'j'}^J$ quantities over all the values of $\Ecol$, $j$, $j'$, and $J$ is equal to 40.4\%, 36.7\%, 27.1\%, and 7.2\% for $v'=0$, $1$, $2$, and $3$, respectively. Thus, partial wave interference in HF$(v')$ forward scattering is increasingly constructive for the larger $v'$ (for details, see the Russian paper by Azriel et al.\ \cite{AAK09}). An almost completely constructive character of the partial wave interference for the HF$(v'=3)$ scattering in the forward direction was first found by Castillo et al.\ \cite{CMS96}, but their analysis was limited to the rotationally unresolved angular distributions and was not carried over to values of $v'\leq 2$ \cite{CMS96}.

\bigskip\medskip

\noindent
{\Large\textbf{IV. Conclusions}}

\bigskip

As we have seen, many features of the rotationally resolved DCSs $\rd\sigma_{v'j'}/\rd\Omega$ for the HF$(v',j')$ product from the $\F+\H_2$ reaction change monotonically as the vibrational quantum number $v'$ grows from $0$ to $3$ (and it is one of the main goals of the paper to attract attention to this phenomenon). For instance, the forward scattering peak increases (section~IIIa), the forward scattering cools down rotationally (section~IIIb), the QM oscillations of the angular distributions become more pronounced (section~IIIc), and the partial wave interference in the small $\theta$ region becomes increasingly constructive (section~IIId). Undoubtedly, these trends also hold for the $\F+\D_2$ reaction. Since they affect \emph{all} the vibrational states of the HF product they seem to hardly arise from a QM resonance. On the other hand, it is the combination of such effects and the energy limitation on the formation of HF$(v'=3)$ molecules with large $j'$ values that favors the forward scattering peak of the HF$(v'=3)$ products. Thus, most probably, QM resonances do not play a key role in the origin of this peak, in agreement with some previous papers \cite{WDQ08,CMS96}.

This conclusion (arrived at without handling resonances themselves) is confirmed by some other facts. The forward scattering peak is observed in the vibrationally resolved angular distributions of the product molecules with the maximal possible $v'$ value from \emph{both} the $\F+\H_2$ reaction and the isotopically substituted $\F+\D_2$ reaction \cite{NWR85a,NWR85b}. It exists in a rather wide range of the collision energies $\Ecol$ for any value $j=0$, $1$, and $2$ of the rotational quantum number of the diatomic reactant. Moreover, as was mentioned in section~I, this vibrationally selective forward peak can be reproduced for both the reactions in quasiclassical trajectory calculations \cite{FMR98,TS88AHN93,ABH94a,ABH94bABH94cABH94dABF96,CHW98,MAB99}. Note that trajectory simulations fail to reproduce the resonance patterns of the $\F+\H\D \to \D+\H\F$ reaction \cite{Liu01CS02,DLL00,Yang07,SSM00aSSM00b,LSM02}.

We believe that our approach based on a careful examination of the state-to-state DCSs for \emph{all} the values of $v'$ and $j'$ could be useful in other situations as well, in particular, in studies of indisputable resonance effects \cite{YZ08,DLL00,QRC06,RCQ06,SSM00aSSM00b,LSM02}.

It is interesting to note that in the very recent paper by Xiahou and Connor \cite{XC09} (which also uses the SW surface), the helicity-resolved DCSs of the $\F+\H_2(v=j=0) \to \H+\H\F(v'=j'=3)$ reaction at a collision energy of 2.74 kcal/mol are found to be an example of broad (attractive) rainbow scattering.

A remaining important task is to clarify the relationship between the features of the PES topography and the dynamical characteristics of the $\F+\H_2$ reactive scattering discussed in this work, for instance, the dominance of the positive forward scattering coefficients $I_{v'j'}$ over the negative ones and their growth as $v'$ increases for fixed small $j'$ values \cite{ARS98aARS98b,RST00RST04}. The correlation techniques we introduced recently \cite{RST00RST04,RST02} could probably be of help here. However, the relative trends in the DCS behavior studied in this work cannot be highly sensitive to the details of the surface used. In particular, all these trends are expected to also hold for the more advanced ground state $\F\H_2$ PESs \cite{CHW98,QRC06,FAC08,HW97,ACF02,ACF03ACF05b,HGM05,XXZ06,FXZ08} listed in section~II.

\bigskip\medskip

\noindent
{\Large\textbf{Acknowledgements}}

\bigskip

The authors are grateful to Prof.\ D.~E.~Manolopoulos for his advice in using the ABC program \cite{SCM00} as well as to D.~B.~Kabanov and E.~V.~Kolesnikova for their help in its installation. The work was partially supported by the Russian Fund for Basic Research (Grant No.\ \mbox{06-03-32159}).

\newpage

{\Large\textbf{Figure Captions}}

\bigskip

\textbf{Figure~1.} Some vibrationally and rotationally resolved DCSs $\rd\sigma_{v'j'}/\rd\Omega$ of the $\F+\H_2(v=0;j=2)$ reaction at the maximum collision energy $\Ecol=3.42$ kcal/mol.

\medskip

\textbf{Figure~2.} The forward scattering coefficients $I_{v'j'}$ of eq~\ref{eq1}.

\medskip

\textbf{Figure~3.} The modified forward scattering coefficients $\Inew_{v'j'}$ of eq~\ref{eq2}.

\medskip

\textbf{Figure~4.} The contributions $D_{v'j'}$ of eq~\ref{eq3} from the rotational levels $j'$ to forward scattering of the HF$(v')$ products. The meaning of the curves is the same as in the $v'=1$ panels of Figures~2 and~3.

\medskip

\textbf{Figure~5.} The relative oscillation amplitudes $A_{v'j'}$ of eq~\ref{eq4}. The meaning of the curves is the same as in the $v'=1$ panels of Figures~2 and~3.

\newpage
\vspace*{-3.5cm}\leavevmode\hspace{-25mm}\includegraphics{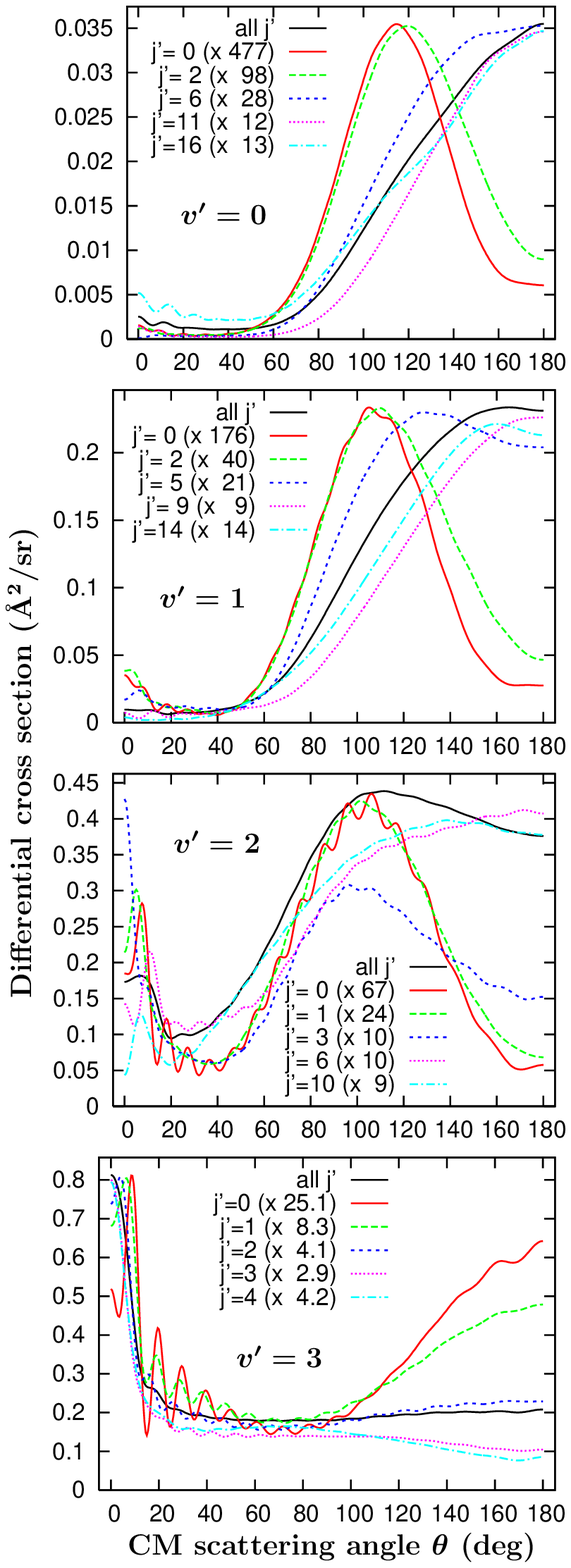}

\vspace*{-5.3cm}\begin{flushright}\large\textbf{Figure~1}\end{flushright}

\newpage
\vspace*{-3.5cm}\leavevmode\hspace{-25mm}\includegraphics{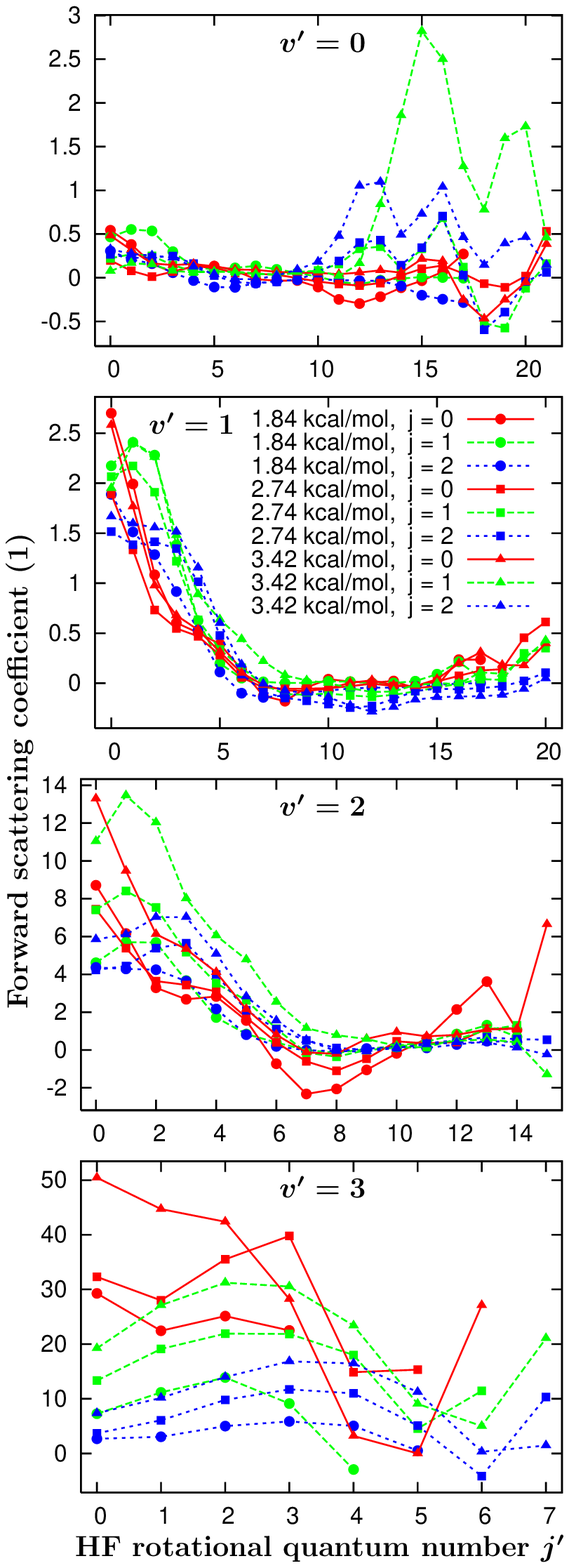}

\vspace*{-5.3cm}\begin{flushright}\large\textbf{Figure~2}\end{flushright}

\newpage
\vspace*{-3.5cm}\leavevmode\hspace{-25mm}\includegraphics{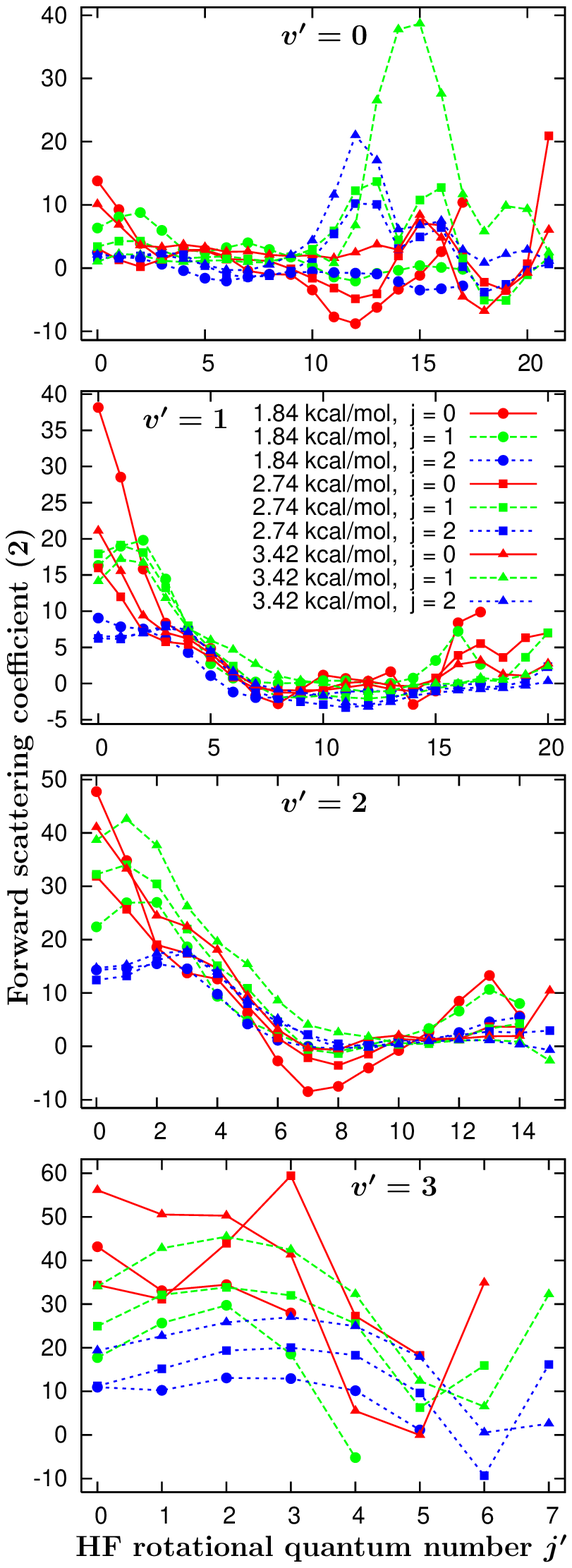}

\vspace*{-5.3cm}\begin{flushright}\large\textbf{Figure~3}\end{flushright}

\newpage
\vspace*{-3.5cm}\leavevmode\hspace{-25mm}\includegraphics{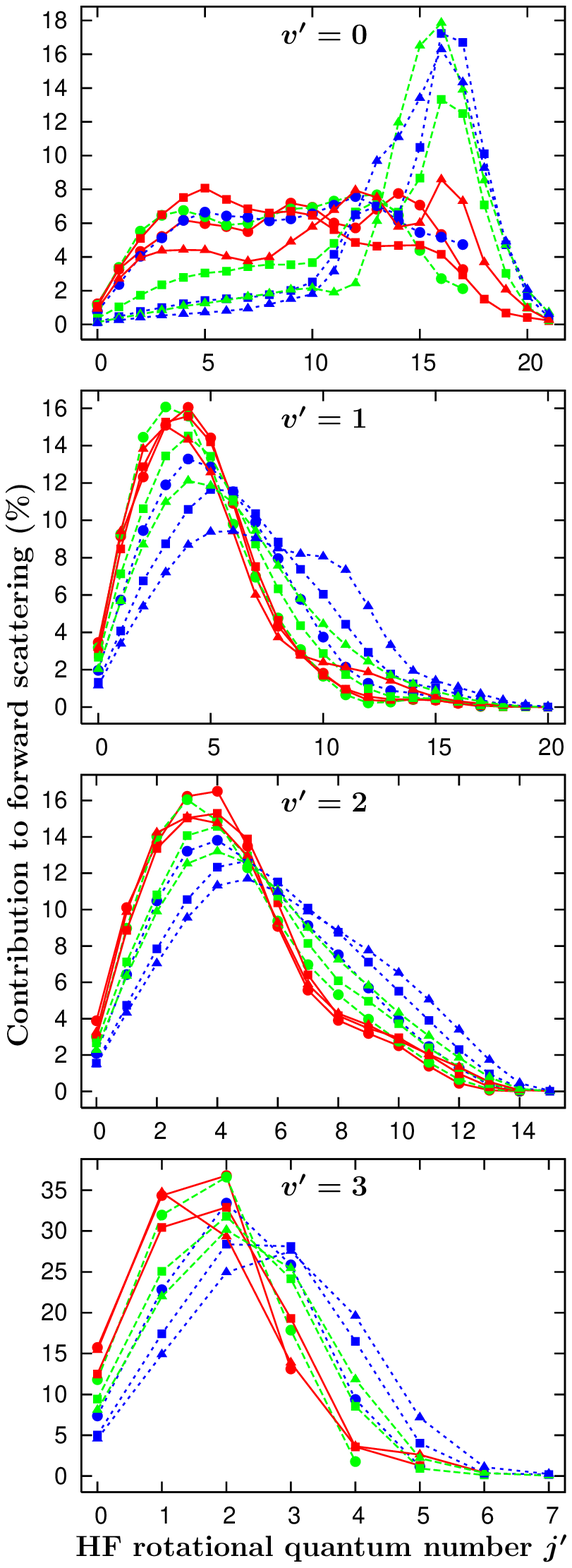}

\vspace*{-5.3cm}\begin{flushright}\large\textbf{Figure~4}\end{flushright}

\newpage
\vspace*{-3.5cm}\leavevmode\hspace{-25mm}\includegraphics{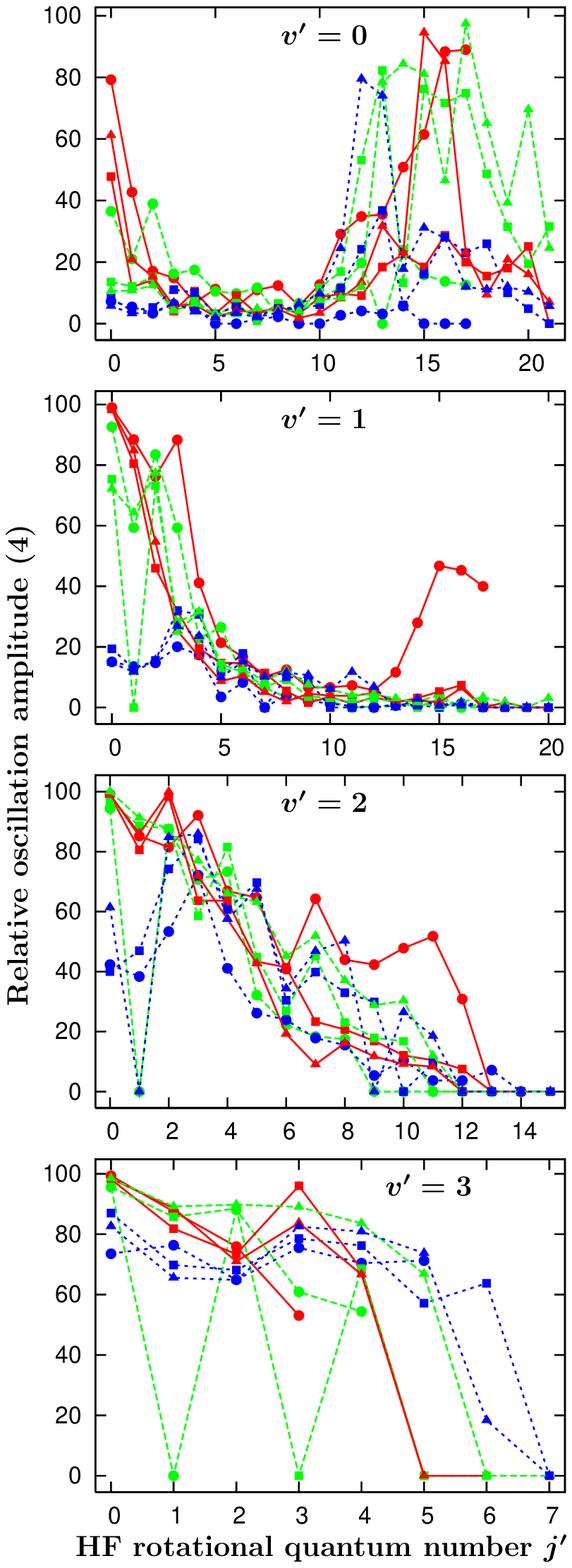}

\vspace*{-5.3cm}\begin{flushright}\large\textbf{Figure~5}\end{flushright}


\begin{thebibliography}{99}
\bibitem{bibnote}
The HF forward scattering direction, which corresponds to the zero of the CM scattering angle $\theta$, is defined to coincide with the fluorine atom beam direction.
\end{thebibliography}
\end{document}